\documentclass[journal]{IEEEtran}
\usepackage{amsfonts}
\IEEEoverridecommandlockouts

\ifCLASSINFOpdf
\else
\fi
\usepackage{epstopdf}
\usepackage{multirow}
\usepackage{cite}
\usepackage{stfloats}
\usepackage{color}
\usepackage{epsfig}
\usepackage{graphicx}  
\usepackage{float}  
\usepackage{psfig}
\usepackage{epsf}
\usepackage{slashbox}
\usepackage{amssymb }
\usepackage[cmex10]{amsmath}
\usepackage{algorithm} 
\usepackage{algorithmic}
\usepackage{booktabs}
\usepackage{amsmath,bm}
\usepackage{colortbl}
\usepackage{caption}
\usepackage{makecell}
\usepackage{subcaption}
\usepackage{adjustbox}

\usepackage[percent]{overpic}
\begin{document}
\title{Gaussian Copula-Based Outage Performance Analysis of Fluid Antenna Systems: Channel Coefficient- or Envelope-Level Correlation Matrix?
}
\author{Rui Xu, Yinghui Ye, Xiaoli Chu, Guangyue Lu, Farshad Rostami Ghadi, and Kai-Kit Wong \vspace{-20pt}
\thanks{Rui Xu, Yinghui Ye, and Guangyue Lu are with the Shaanxi Key Laboratory of
Information Communication Network and Security, Xi'an University of Posts
\& Telecommunications, China (e-mails: dhlscxr@126.com; connectyyh@126.com; tonylugy@163.com). }
\thanks{Xiaoli Chu is with the School of Electrical and Electronic Engineering, University of Sheffield, Sheffield, U.K. (e-mail: x.chu@sheffield.ac.uk).}
\thanks{Farshad Rostami Ghadi and Kai-Kit Wong are with the Department of Electronic and Electrical Engineering, University College London, WC1E 7JE London, U.K., and Kai-Kit Wong is also with
the Yonsei Frontier Laboratory, Yonsei University, Seoul 03722, South Korea
(e-mails: f.rostamighadi@ucl.ac.uk; kai-kit.wong@ucl.ac.uk).}
}
\markboth{}
{Shi\MakeLowercase{\textit{et al.}}:}
\maketitle
\begin{abstract}

Gaussian copula has been employed to evaluate the outage performance of Fluid Antenna Systems (FAS), with the covariance matrix reflecting the dependence among multivariate normal random variables (RVs).
While prior studies approximate this matrix using the channel coefficient correlation matrix from Jake's model, this work instead employs the channel envelope correlation matrix, motivated by the fact that the multivariate normal RVs are generated by transforming correlated channel envelopes. This raises an open question of whether using the coefficient- or envelope-level correlation matrix yields better accuracy in accessing FAS performance.
Toward this end, this paper explores the benefits of using the envelope-level correlation matrix under fully correlated Nakagami-$m$ fading, and develops a method for generating such fading channels for Monte Carlo simulations, which serve as a benchmark for validating the theoretical results.
Simulation results confirm the effectiveness of the proposed channel modeling approach and demonstrate the superior accuracy of using the envelope-level correlation matrix, particularly in sparse port deployment and low-outage regime.
\end{abstract}
\vspace{-5pt}
\IEEEpeerreviewmaketitle
\vspace{-8pt}
\section{Introduction}

\IEEEPARstart{R}{ecently}, fluid antenna (FA) has garnered increasing attention in wireless communication communities \cite{10678877, 10753482}.
Its key idea is to densely deploy ports within a predefined space and dynamically move its RF front-end to the port with the best channel condition for receiving/transmitting signals, thereby increasing spatial diversity.
Accordingly, compared to the traditional
fixed antenna systems (TAS), FA systems (FAS) enjoy a higher transmission reliability while
using much fewer antennas and RF front-ends.
However, closely packed ports unavoidably increase spatial correlation among channels, which poses significant challenges for FAS performance evaluation.

A rigorous theoretical assessment of FAS performance requires the spatial correlation model to accurately reflect the propagation conditions encountered in practical FAS deployments. To this end, various correlation models have been introduced, including the reference port model in \cite{9264694}, the equally correlated model in \cite{wong2022closed}, the block-diagonal correlation model in \cite{10623405}, and the fully correlated model in \cite{10103838}.
Among these models, the fully correlated model in \cite{10103838} most accurately captures the spatial correlation, where the channel at each port is modeled as a weighted linear combination of the channels at all the other ports, with the weights determined by the eigenvalues and eigenvectors of the correlation matrix.
Although the two-stage approximation proposed in \cite{10103838} results in a simplified OP expression, the analytical framework is not easily extendable to the generalized Nakagami-$m$ fading.
Meanwhile, we note that in copula theory \cite{9159617}, the multivariate distribution of correlated random
variables (RVs) can be generated from the marginal distributions under arbitrary fading, and that the dependency structure can be modeled using the copula function. As a type of elliptical copula, Gaussian copula has attracted considerable attention in the performance analysis of FAS \cite{10678877}, since unlike other types of copulas such as Archimedean
copulas, it incorporates a covariance matrix to capture the dependence among multivariate normal RVs.
In existing studies, the covariance matrix was approximated in \cite{10678877} using the channel coefficient correlation matrix from Jake's model. Accordingly, Gaussian copula links the FAS performance to the number of ports and size of FA, enabling a more accurate modeling under different port deployment densities compared to other types of copula.

Most studies on FAS performance \cite{9264694,wong2022closed,10623405,10103838} have been limited to Rayleigh fading, which only applies to non-line-of-sight and scattering-rich propagation environments. To fill this gap, the authors in \cite{10678877} pioneered the outage probability (OP) analysis of FAS under Nakagami-$m$ fading through Gaussian copula.
Taking detailed consideration, two issues remain to be addressed.
First, the covariance matrix in Gaussian copula function reflects the dependence among multivariate normal RVs transformed from correlated channel envelopes, and Gaussian copula is employed to describe the joint distribution of multiple correlated channel envelopes \cite{10678877}.
While existing studies approximate this covariance matrix using the coefficient-level correlation matrix from Jake's model, an open question arises: does using the envelope-level correlation matrix yield more accurate performance than the coefficient-level one\footnote{{Notably, \cite{10678877} proposed an innovative theoretical framework for characterizing the outage performance of FAS using Gaussian copula. Building upon this framework, we approximate the covariance matrix using the envelope-level correlation matrix rather than coefficient-level correlation matrix, and examine its potential improvements in modeling accuracy.}}?
Second, accurately generating fully correlated Nakagami-$m$ fading channel is essential for accurate evaluation of FAS performance via Monte Carlo simulation to validate the accuracy of theoretical results. Although \cite{1356206} have proposed a method for generating correlated Nakagami-$m$ fading channel under the equally correlated model, a fully correlated Nakagami-$m$ fading channel generation method is still lacking.

Motivated by these observations, this paper explores the benefits of using the envelope-level correlation matrix under fully correlated Nakagami-$m$ fading.
To validate the modeling accuracy of Gaussian copula with different correlation matrices, we develop a method for generating fully correlated Nakagami-$m$ fading channels, which serves as the foundation for obtaining accurate OP through Monte Carlo simulations as a benchmark.
Simulation results demonstrate the effectiveness of the proposed channel construction approach, and validate the superiority of Gaussian copula using the envelope-level correlation matrix over coefficient-level one, particularly in sparse port deployment and low-outage regime.
\vspace{-6pt}

\section{Gaussian Copula-based Approach for FAS}
We consider a point-to-point FAS, where a single-fixed-antenna transmitter sends information to a receiver equipped with a FA. The FA contains ${N}$ preset ports that are equally distributed along a straight line of length ${W}\lambda$, where $\lambda$ is the wavelength of the carrier signal and $W$ denotes the size of FA.
The OP of FAS depends on the distribution of the peak channel envelope, which is defined as $\left| {{h_{{\rm{FAS}}}}} \right| = \max \left\{ {\left| {{h^{\left( 1 \right)}}} \right|, \cdots ,\left| {{h^{\left( N \right)}}} \right|} \right\}$, where $\left| {{h^{\left( n \right)}}} \right|$, $n \in {\cal N} = \left\{ {1, \cdots ,N} \right\}$, denotes the channel envelope at the $n$-th port.
To obtain the distribution of $\left| {{h_{\rm{FAS}}}} \right|$,
Gaussian copula was used in \cite{10678877} to derive the CDF of $\left| {{h_{{\rm{FAS}}}}} \right|$, through linking the CDFs of ${\left| {{h^{\left( n \right)}}} \right|}$ and the CDF of $\left| {{h_{{\rm{FAS}}}}} \right|$ via a Gaussian copula function with a covariance matrix, which is provided in \textbf{Theorem 1}.
%

\textbf{Theorem 1:} The CDF of $\left| {{h_{{\rm{FAS}}}}} \right|$ under fully correlated channel model can be derived as
 \begin{align}\label{1A}
{F_{\left| {{h_{{\rm{FAS}}}}} \right|}}\!\left( r \right) \!\!=\!\! {\Phi _{{{\bf{R}}}}}\!\!\left( \!{{\phi ^{ - 1}}\!\!\left( \! {{F_{\left| {{h^{\left( 1 \right)}}} \right|}}\!\!\left( r \right)}\! \right)\!, \ldots ,{\phi ^{ - 1}}\!\!\left( \!{{F_{\left| {{h^{\left( N \right)}}} \right|}}\!\!\left( r \right)} \!\right)} \!\right),
\end{align}
where ${F_{\left| {{h^{\left( n \right)}}} \right|}}\left( r \right)$ is the CDF of $\left| {{h^{\left( n \right)}}} \right|$, ${\phi ^{ - 1}}\left( {{F_{\left| {{h^{\left( n \right)}}} \right|}}\left( r \right)} \right) = \sqrt 2 {{\mathop{\rm erf}\nolimits} ^{ - 1}}\left( {2{F_{\left| {{h^{\left( n \right)}}} \right|}}\left( r \right) - 1} \right)$, ${\phi ^{ - 1}}\left(  \cdot  \right)$ is the inverse CDF of a standard normal distribution, ${{\mathop{\rm erf}\nolimits} ^{ - 1}}\left(  \cdot  \right)$ is the inverse error function, ${\Phi _{{{\bf{R}}}}}\left(  \cdot  \right)$ is the joint CDF of the multivariate normal distribution with a zero mean vector,
and the covariance matrix ${\bf{R}}$ of multivariate normal RVs.


\emph{Remark 1:} \emph{In Gaussian copula, the covariance matrix, which reflects the correlation characteristic of multivariate normal RVs, was approximated by the channel coefficient correlation matrix ${\bf{J}}$ from Jake's model in \cite{10678877}. However, the normal RV ${\phi ^{ - 1}}\left( {{F_{\left| {{h^{\left( n \right)}}} \right|}}\left( r \right)} \right)$ is transformed from the
RV $r$ of the channel envelope $\left| {{h^{\left( n \right)}}} \right|$, and the Gaussian copula function in \eqref{1A} describes the distribution of correlated channel envelopes.
Accordingly, it remains a worthwhile question whether approximating the covariance matrix in Gaussian copula using the envelope-level rather than coefficient-level correlation matrix yields more accurate characterization of FAS performance.}
Note that the coefficient-level correlation matrix ${\bf{J}}$ is obtained from Jake's model, where each entry ${{\bf{J}}_{n,\tilde n}} = {J_0}\left( {{{2\pi \left( {n - \tilde n} \right)W} \mathord{\left/
 {\vphantom {{2\pi \left( {n - \tilde n} \right)W} {\left( {N - 1} \right)}}} \right.
 \kern-\nulldelimiterspace} {\left( {N - 1} \right)}}} \right)$ denotes the spatial correlation between the $n$-th port and the ${\tilde n}$-th port, with ${J_0}\left(  \cdot  \right)$ denoting the zero-order Bessel function of the first kind, and the envelope-level correlation matrix ${\bf{J}_h}$ is derived from ${\bf{J}}$ after some manipulations, as shown in \eqref{7A} in the next section.

\begin{figure}[htbp]
  \centering
  \begin{subfigure}[t]{0.23\linewidth}
    \includegraphics[width=\linewidth]{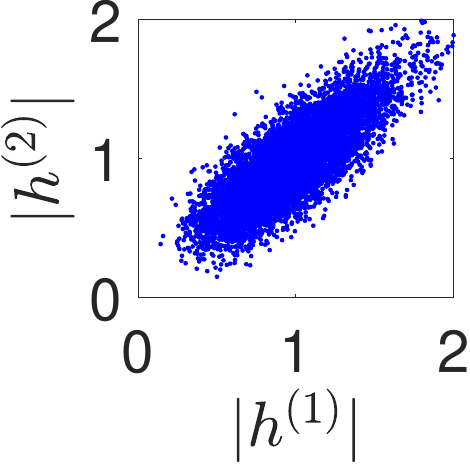}
    \caption{$W=0.1$}
  \end{subfigure}
  \begin{subfigure}[t]{0.23\linewidth}
    \includegraphics[width=\linewidth]{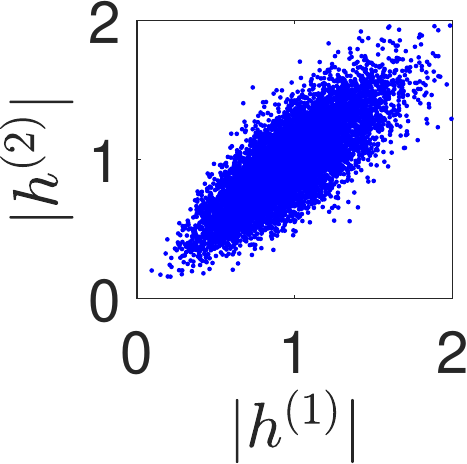}
    \caption{$W=0.1$}
  \end{subfigure}
  \begin{subfigure}[t]{0.23\linewidth}
    \includegraphics[width=\linewidth]{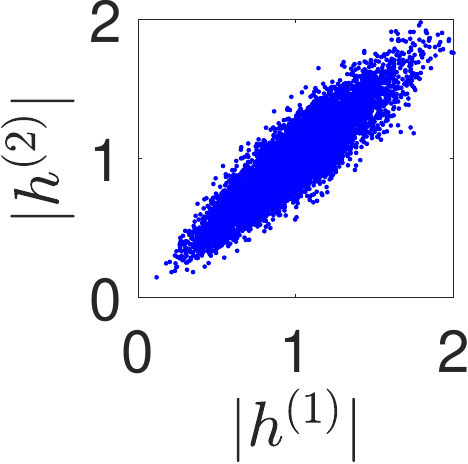}
    \caption{$W=0.1$}
  \end{subfigure}
  \begin{subfigure}[t]{0.23\linewidth}
    \includegraphics[width=\linewidth]{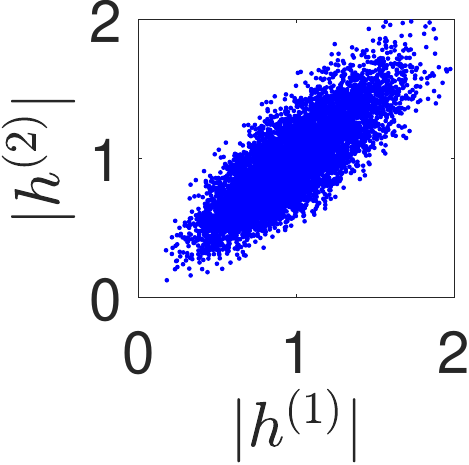}
    \caption{$W=0.1$}
  \end{subfigure}

  \vspace{2mm}

  \begin{subfigure}[t]{0.23\linewidth}
    \includegraphics[width=\linewidth]{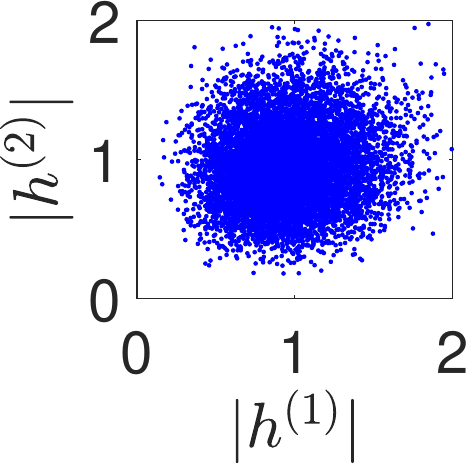}
    \caption{$W=0.3$}
  \end{subfigure}
  \begin{subfigure}[t]{0.23\linewidth}
    \includegraphics[width=\linewidth]{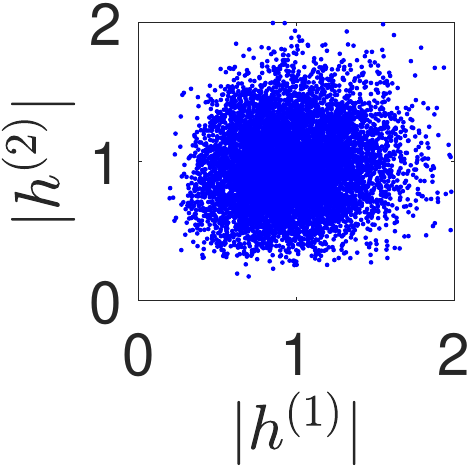}
    \caption{$W=0.3$}
  \end{subfigure}
  \begin{subfigure}[t]{0.23\linewidth}
    \includegraphics[width=\linewidth]{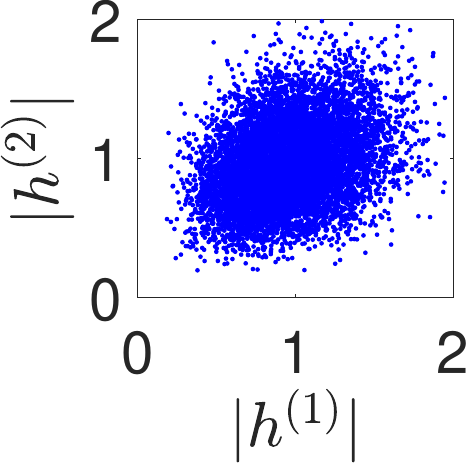}
    \caption{$W=0.3$}
  \end{subfigure}
  \begin{subfigure}[t]{0.23\linewidth}
    \includegraphics[width=\linewidth]{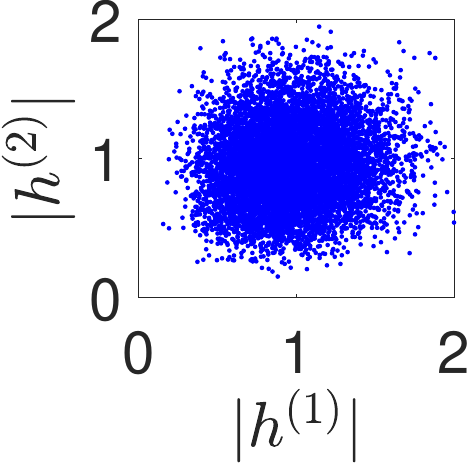}
    \caption{$W=0.3$}
  \end{subfigure}

  \vspace{2mm}

  \begin{subfigure}[t]{0.23\linewidth}
    \includegraphics[width=\linewidth]{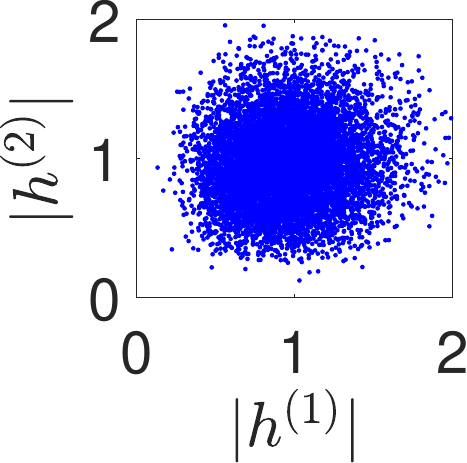}
    \caption{$W=0.5$}
  \end{subfigure}
  \begin{subfigure}[t]{0.23\linewidth}
    \includegraphics[width=\linewidth]{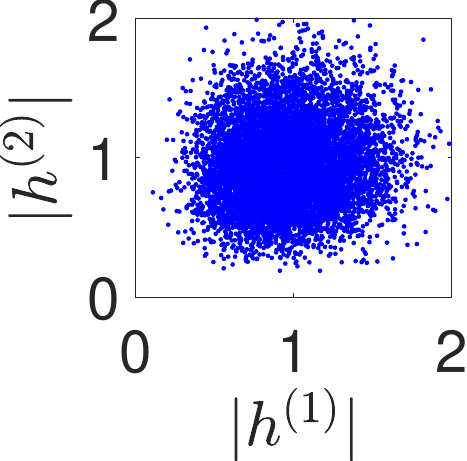}
    \caption{$W=0.5$}
  \end{subfigure}
  \begin{subfigure}[t]{0.23\linewidth}
    \includegraphics[width=\linewidth]{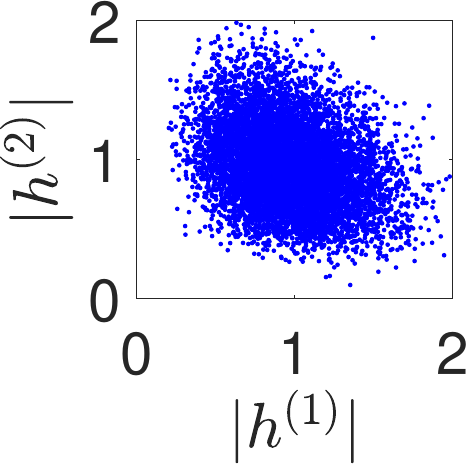}
    \caption{$W=0.5$}
  \end{subfigure}
  \begin{subfigure}[t]{0.23\linewidth}
    \includegraphics[width=\linewidth]{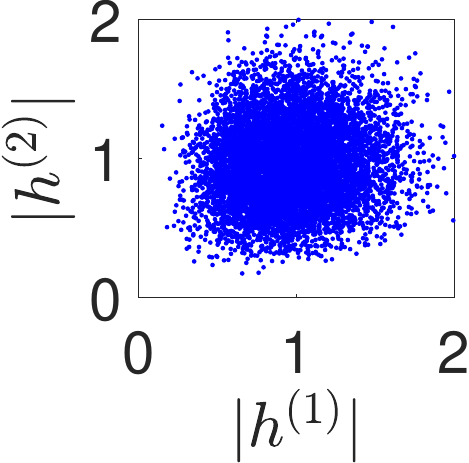}
    \caption{$W=0.5$}
  \end{subfigure}

  \caption{\small{The scatterplots of envelopes following Nakagami-$m$ fading.
}}
\vspace{-1.8em}
  \label{fig:3x4}
\end{figure}

Next, we compare the modeling accuracy of Gaussian copula when using coefficient- and envelope-level correlation matrices under fully correlated Nakagami-$m$ fading. Fig. 1 compares the scatterplots of the channel envelopes ${\left| {{h^{\left( 1 \right)}}} \right|}$ and ${\left| {{h^{\left( 2 \right)}}} \right|}$ following Nakagami-$m$ fading obtained from simulations (i.e., (a), (e), (i)), and Gaussian copula with the correlation matrix ${\bf{R}}$ of multivariate normal RVs\footnote{The correlation matrix ${\bf{R}}$ can be measured as follows. We first generate the Nakagami-$m$ distributed samples and compute their CDF, which follows the uniform distribution. Then, the uniformly distributed samples are mapped as the normally distributed samples through the inverse Gaussian conversion, and then their correlation is calculated.} (i.e., (b), (f), (j)), the coefficient-level correlation matrix ${\bf{J}}$ (i.e., (c), (g), (k)), or the envelope-level correlation matrix ${\bf{J}_h}$ (i.e., (d), (h), (l)).\footnote{Under different correlation matrices, multidimensional uniformly distributed samples are generated from a Gaussian copula, and then the envelope samples of $\left| {{h^{\left( n \right)}}} \right|$ can be measured by applying the inverse CDF of Nakagami-$m$ distribution to the uniformly distributed samples.}
It can be observed that the scatterplots of the channel envelopes under Gaussian copula with the envelope-level correlation matrix and the correlation matrix of multivariate normal RVs are closer to those obtained from simulations, compared to that with the coefficient-level correlation matrix.
Specifically, when the port density is high, e.g., $W=0.1$ and $N=2$, the correlation between the channel envelopes generated based on the coefficient-level correlation matrix is stronger than that observed in simulations. This leads to an underestimation of the spatial diversity of FAS and may overestimate the OP.
Conversely, when the port density is low, e.g.,
$W=0.5$ and $N=2$, the channel envelopes generated using the coefficient-level correlation matrix exhibit stronger negative correlation compared to simulations. This results in fewer samples in the lower-left tail of the distribution, which is critical in determining outage events, and may lead to an underestimation of OP.
Moreover, although channel envelopes generated using either the correlation matrix of multivariate normal RVs or that of the channel envelopes can better capture the actual sample correlation, little discrepancies still exist particularly at the lower-left tail when the port density is high, potentially introducing performance errors.
Such small gap between theory and simulation is expected since the Gaussian copula-based method is an approximation.
Furthermore, Fig. 2 compares the PDFs and CDFs of $\left| {{h_{{\rm{FAS}}}}} \right|$ under $N=2$, obtained from \eqref{1A} using coefficient- and envelope-level correlation matrices respectively, as well as from Monte-Carlo simulation. In Fig. 2(a), Gaussian copula with envelope-level correlation matrix provides more accurate channel distribution compared to that with coefficient-level one.
Moreover, since practical outage requirements are usually below ${\rm{1}}{{\rm{0}}^{ - 2}}$, i.e., within the CDF value less than ${\rm{1}}{{\rm{0}}^{ - 2}}$, Fig. 2(b) shows that Gaussian copula with envelope-level correlation matrix can yields a more accurate performance assessment in this regime.

\begin{figure}
\centering
\begin{subfigure}[t]{0.49\linewidth}
  \centering
  \includegraphics[width=\linewidth]{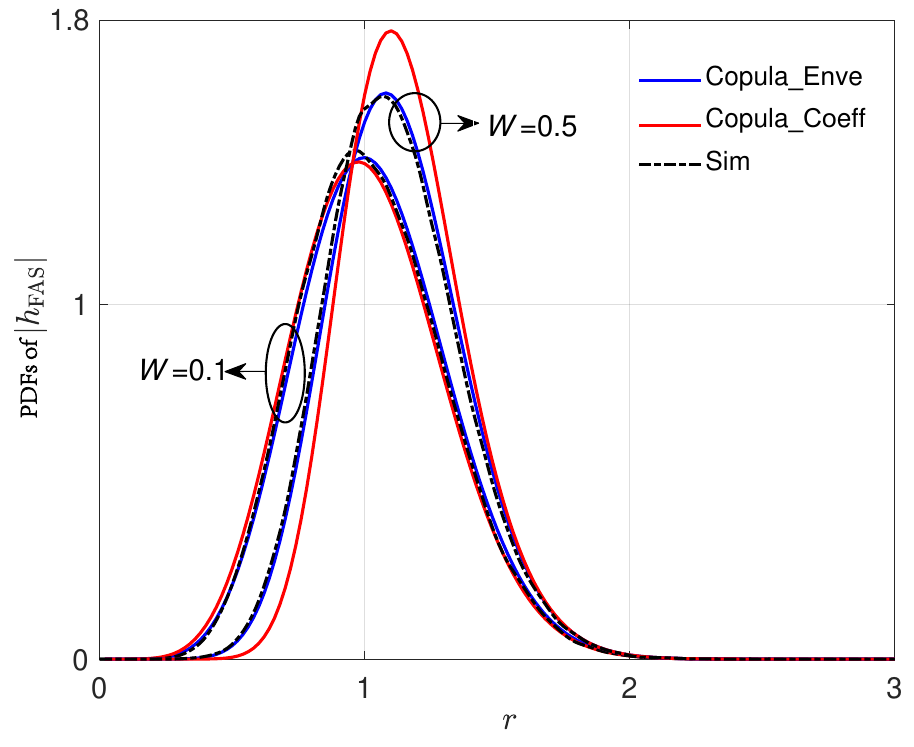}
  \vspace{-1.5em}
  \caption{\small{The PDFs of $\left| {{h_{{\rm{FAS}}}}} \right|$.}}
  \label{fig:subfig:a}
\end{subfigure}
\hfill
\begin{subfigure}[t]{0.49\linewidth}
  \centering
  \includegraphics[width=\linewidth]{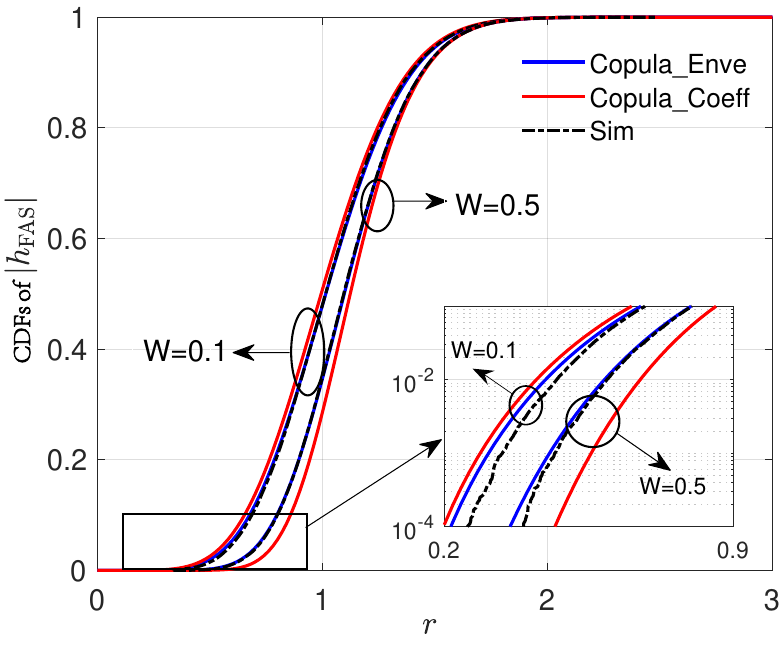}
  \vspace{-1.5em}
  \caption{\small{The CDFs of $\left| {{h_{{\rm{FAS}}}}} \right|$.}}
  \label{fig:subfig:b}
\end{subfigure}

\vspace{-0.5em}
\caption{\small{The PDFs and CDFs of $\left| {{h_{{\rm{FAS}}}}} \right|$ under simulation and Gaussian copula with different correlation matrices when $N=2$,$m=3$,$\mu=1$.}}
\vspace{-1.8em}
\label{fig:subfig}
\end{figure}

\vspace{-5pt}
\section{Construction of Fully Correlated Nakagami-$m$ Fading Channel in FAS}
%

This section presents a method for generating fully correlated Nakagami-$m$ fading channel envelopes to enable accurate performance evaluation via Monte-Carlo simulation, which serves as a benchmark to assess the modeling accuracy of Gaussian copula using different correlation matrices.

When generating fully correlated Nakagami-$m$ fading channel envelopes, the following two conditions must be satisfied:

 (i) The generated channel envelopes follow Nakagami-$m$ distribution with shape parameter $m$ and scale parameter $\mu$.

 (ii) According to \cite{1356206},
the correlation matrix of the generated channel envelopes ${{\bf{J_h}}}$ should satisfy
\begin{align}\label{7A}
{{\bf{J}}_{\bf{h}}}^{\left( {n,\tilde n} \right)} = \frac{{F\left( { - \frac{1}{2}, - \frac{1}{2};m;{{\bf{J}}_{\bf{r}}}^{\left( {n,\tilde n} \right)}} \right) - 1}}{{\psi (m) - 1}}, \tilde n \in {\cal N}, n \ne \tilde n,
 \end{align}
 where $\psi (m) = \Gamma (m)\Gamma (m + 1)/{\Gamma ^2}(m + {1 \mathord{\left/
 {\vphantom {1 2}} \right.
 \kern-\nulldelimiterspace} 2})$, $F ( \cdot )$ is the hypergeometric function, and ${{{\bf{J}}_{\bf{r}}}}$ denotes the channel gain correlation matrix. Meanwhile, the channel gain correlation matrix ${{{\bf{J}}_{\bf{r}}}}$ should satisfy ${{\bf{J}}_{\bf{r}}}^{\left( {n,\tilde n} \right)} \!=\! {\left( {{{\bf{J}}^{\left( {n,\tilde n} \right)}}} \right)^2}$, as shown in \cite{stuber2001principles}.

Toward this end, when the fading severity index $m$ is an integer, the parameterization of correlated Nakagami-$m$ fading channel envelopes at the $n$-th port can be achieved by using the square root of the sum of $m$ squared Rayleigh RVs, i.e.,
\begin{align}\label{3A}
\left| {{h^{\left( n \right)}}} \right| = \sqrt {{R_n}}  = \sqrt {\sum\nolimits_{j = 1}^m {{{\left| {{{\bf{G}}_{nj}}} \right|}^2}} },
 \end{align}
where $n \in {\cal N}$, $j \in {\cal M} = \left\{ {1, \cdots ,m} \right\}$, ${R_n} = \sum\nolimits_{j = 1}^m {{{\left| {{{\bf{G}}_{nj}}} \right|}^2}} $ denotes the channel gain.

To satisfy the correlation of channel coefficients, we perform eigenvalue decomposition on the coefficient-level correlation matrix ${\bf{J}} \!= \!{\bf{U\Lambda }}{{\bf{U}}^{\mathcal{H}}}$,
where $\mathcal{H}$ stands for conjugate transpose, ${\bf{\Lambda}}\! =\! {\rm{diag}}\left( {{\lambda _1}, \cdots ,{\lambda _N}} \right)$ is the diagonal eigenvalue matrix with ${\lambda _1} \!\!\ge\! \cdots\! \ge\!\! {\lambda _N}$, and $\bf{U}$ is the eigenvector matrix.
 Then, a set of $Nm$ complex Gaussian RVs is generated by\footnote{Different from the equally correlated model that employs a common correlation coefficient \cite{1356206}, a correlation matrix ${\bf{J}}$ is employed under the fully correlated model, which leads to differences in how a set of complex Gaussian RVs is generated for constructing
 $\left| {{h^{\left( n \right)}}} \right|$ to satisfy the expected correlation.}
\begin{align}\label{4A}
{{\bf{G}}_{nj}} = {\bf{U}}{{\bf{\Lambda}} ^{\frac{1}{2}}}{{\bf{Z}}_{nj}}.
 \end{align}
where ${{\bf{Z}}_{nj}} \sim {\cal C}{\cal N}\left( {0,{{\mu \mathord{\left/
 {\vphantom {1 m}} \right.
 \kern-\nulldelimiterspace} m}}} \right)$ for any fixed $n$ and $j$ are independent, and $\mu$ denotes the scale parameter of Nakagami-$m$ distribution.
  It can be easily checked that ${{\bf{G}}_{nj}} \sim {\cal C}{\cal N}\left( {0,{\mu  \mathord{\left/
 {\vphantom {\mu  m}} \right.
 \kern-\nulldelimiterspace} m}} \right)$. For any given $n$, ${{{\bf{G}}_{nj}}}$, $j \in {\cal M}$, are independent,
and thus ${R_n}  \sim {\chi _{2m}}\left( {0,{\mu \mathord{\left/
 {\vphantom {1 2}} \right.
 \kern-\nulldelimiterspace} 2m}} \right)$. After taking the square root of ${R_n}$, $\left| {{h^{\left( n \right)}}} \right|= \sqrt {{R_n}}  \sim {\rm{Nakagami}}\left( {m,\mu} \right)$. Following this process, the two conditions mentioned above can be satisfied.

 \emph{Remark 2:} Typically, the Nakagami-$m$ distribution can be obtained by taking the square root of a Gamma distribution. If the envelope correlation is introduced into the Nakagami-$m$ distribution using a method similar to \eqref{4A}, the resulting distributions will no longer follow Nakagami-$m$ distribution, as it lacks additivity.
Moreover, introducing gain correlation into the Gamma distribution first and then applying the square root transformation to obtain fully correlated Nakagami-$m$ channel envelopes is also infeasible. This is because although Gamma distribution is additive, i.e., it remains Gamma distribution after introducing gain correlation, its parameters are altered. Consequently, the parameters of the resulting Nakagami-$m$ distribution, derived by taking the square root of the correlated Gamma distribution, will also be altered.

Then, we represent the correlation matrix of the generated channel envelopes. The cross-correlation coefficients of ${{\bf{G}}_{nj}}$ and ${{{\bf{G}}_{\tilde ni}}}$ and that of ${R_n}$ and ${R_{\tilde n}}$ are given, respectively, by
 \begin{align}\label{5A}
{\bf{J_g}}^{\!\left( \!{nj,\tilde ni} \!\right)\!} \!\!=\!\! \frac{\!{{\mathbb{E}}\!\left( \!{{{\bf{G}}_{nj}}{\bf{G}}_{\tilde ni}^*} \!\right)}}{{\sqrt {{\mathbb{E}}\!\left( {|{{\bf{G}}_{nj}}{|^2}} \! \right)\!{\mathbb{E}}\!\left(\! {|{{\bf{G}}_{\tilde ni}}{|^2}} \!\right)} }} \!\!=\!\! \left\{ {\begin{array}{*{20}{c}}\!\!\!\!
\!\!\!\!\!{{{\bf{J}}_{n,\tilde n}},\!n\! \ne \tilde \!n\& j \!= \!i},\\
\!\!\!\!{0,j \!\ne \!i},\quad\quad\quad\quad\quad\,
\end{array}} \right.
  \end{align}
\vspace{-10pt}
 \begin{align}\label{6A}
 {\bf{J_r}}^{\left( {n,\tilde n} \right)} = \frac{{{\mathbb{E}}\left( {{R_n}{R_{\tilde n}}} \right)}}{{\sqrt {{\mathbb{E}}\left( {R_n^2} \right){\mathbb{E}}\left( {R_{\tilde n}^2} \right)} }} = {\bf{J}}_{n,\tilde n}^2,n \ne \tilde n.
  \end{align}
 The above relationship between the channel coefficient correlation matrix ${\bf{J}}$ and the channel gain correlation matrix ${{\bf{J_r}}}$, which is consistent with the relationship specified in \cite{stuber2001principles}, verifies the correctness of the proposed channel envelope construction method in \eqref{3A} and \eqref{4A}.
 With obtained ${{\bf{J_r}}}$, the channel envelope correlation matrix ${{\bf{J}}_{\bf{h}}}$ can be obtained through (2).


\begin{figure}
\centering
\begin{subfigure}[t]{0.49\linewidth}
  \centering
  \includegraphics[width=\linewidth]{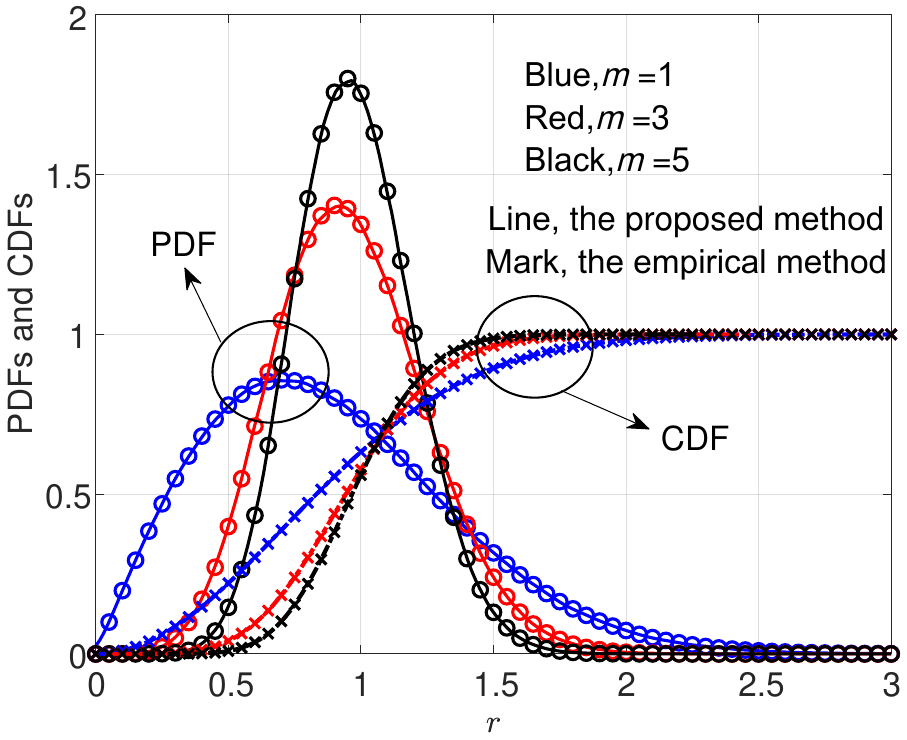}
  \caption{\small{The PDFs and CDFs of Nakagami-$m$ channel envelopes generated by the proposed and the empirical approaches.}}
  \label{fig:subfig:a}
\end{subfigure}
\hfill
\begin{subfigure}[t]{0.49\linewidth}
  \centering
  \includegraphics[width=\linewidth]{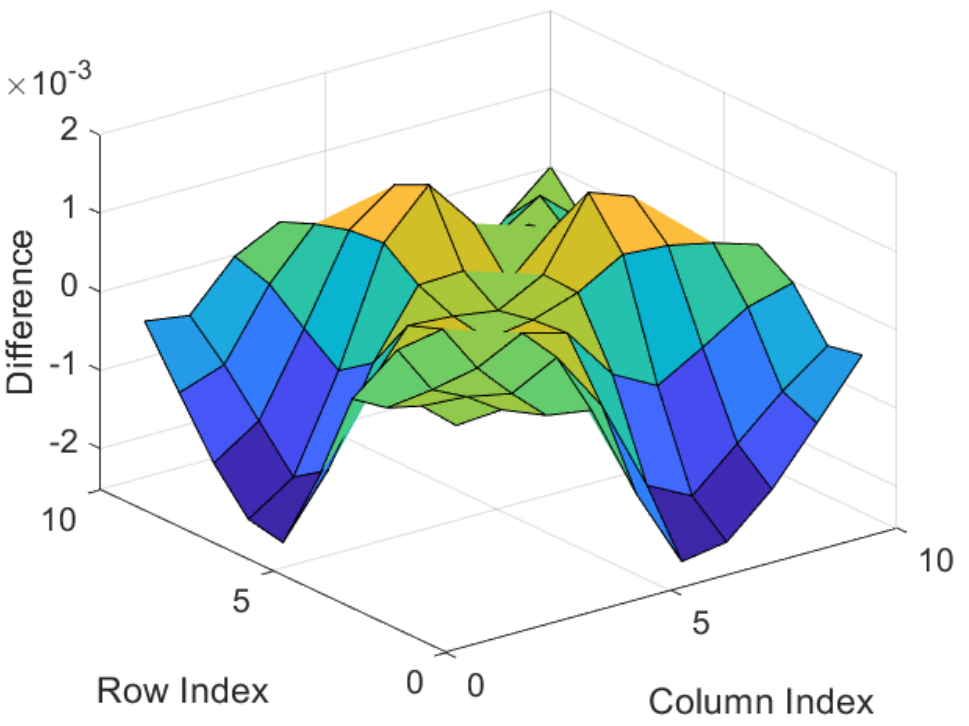}
  \caption{\small{The differences between the correlation matrix of the generated envelopes using the proposed approach and the envelope correlation matrix computed by (2).}}
  \label{fig:subfig:b}
\end{subfigure}

\vspace{-0.5em}
\caption{\small{The distributions and correlation verification of the Nakagami-$m$ channel envelopes generated by the proposed approach.}}
\vspace{-1.8em}
\label{fig:subfig}
\end{figure}


To verify the correctness of the proposed fully correlated Nakagami-$m$ channel envelopes construction approach, simulation results in terms of the distribution of the generated channel envelopes and the correlation among them are provided. Fig. 3 (a) shows the comparison of PDFs and CDFs of the generated channel envelopes using the proposed approach and the empirical approach.
 It can be observed from Fig. 3 (a) that the PDFs and CDFs of the channel envelopes generated by the proposed approach and the empirical approach are perfectly matched. Meanwhile, a quantitative validation based on the root mean square error (RMSE) was conducted, showing that the RMSEs of both the PDFs and CDFs in Fig. 3 (a) under different values of $m$ are all below ${\rm{1}}{{\rm{0}}^{ - 3}}$.
 Moreover, Fig. 3 (b) shows the differences between the correlation matrix of the channel envelopes generated by the proposed approach and
the channel envelope correlation matrix according \eqref{7A}. We can see that the magnitude of differences is below ${10^{ - 2}}$, which is less than $1\% $ with respect to the value of correlation $[-1,1]$.
These satisfy the above mentioned conditions, proving the feasibility of the proposed fully correlated Nakagami-$m$ channel envelopes construction approach.


\section{Numerical Results}

%

\begin{figure*}[htbp]
    \centering
    \begin{minipage}{0.32\linewidth} 
        \centering
        \includegraphics[width=\linewidth]{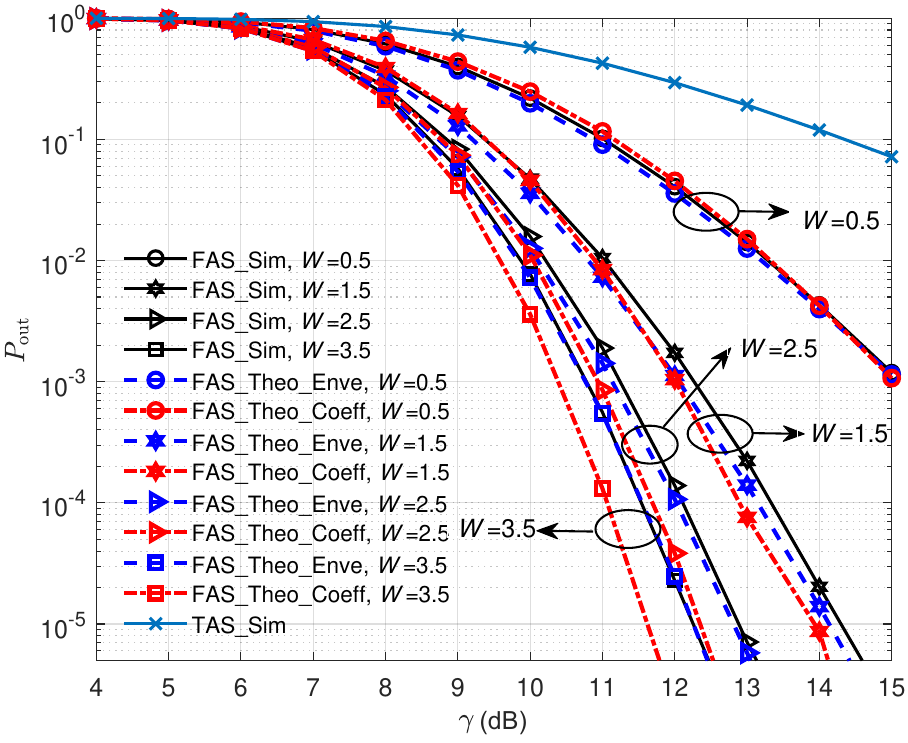}
        \caption{\small{The OPs versus the average transmit SNR $\gamma$ for selected values of $W$ when $m=3$, $\mu=1$, $N=10$, and ${\gamma _{{\rm{th}}}} = 10$ dB.}}
        \label{chutian1}
    \end{minipage}
    \hfill
    \begin{minipage}{0.64\linewidth} 
        \centering
        \begin{subfigure}[b]{0.49\linewidth}
            \centering
            \includegraphics[width=\linewidth]{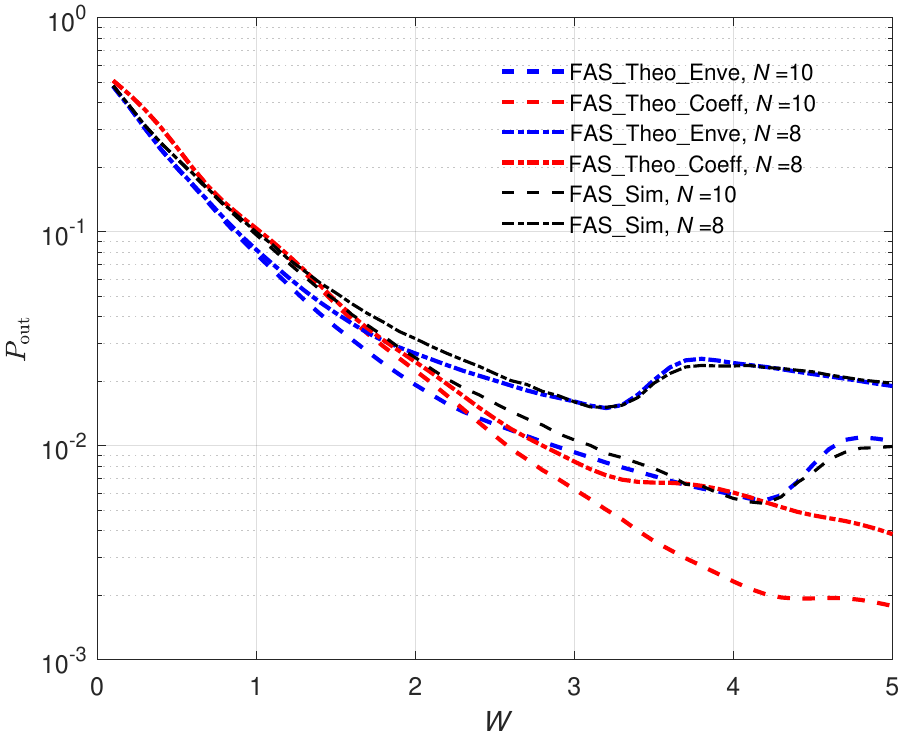}
            \caption{\small{The OPs versus the size of FA $W$ for selected values of $N$ when ${\gamma}=10$ dB.}}
            \label{chutian2}
        \end{subfigure}
        \hfill
        \begin{subfigure}[b]{0.49\linewidth}
            \centering
            \includegraphics[width=\linewidth]{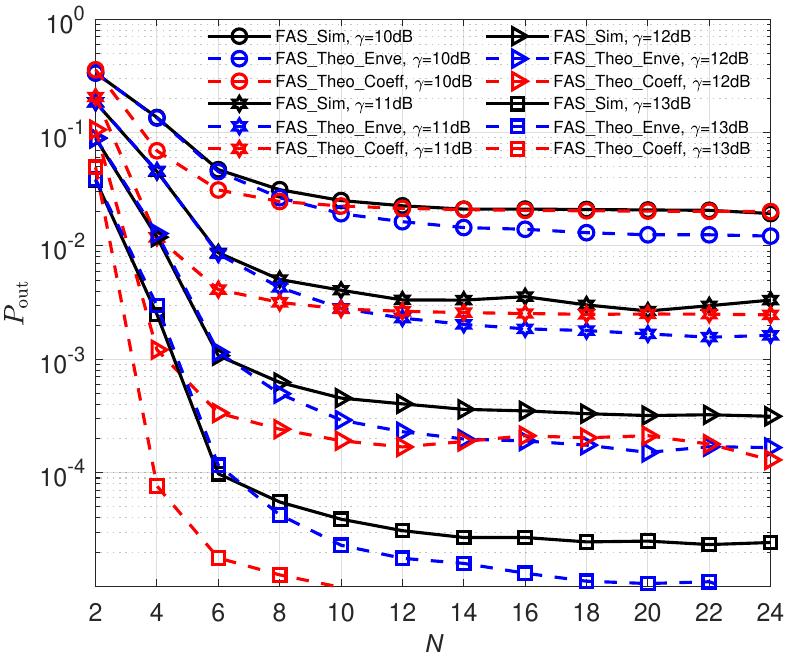}
            \caption{\small{The OPs versus the number of ports $N$ for selected values of $\gamma$ when $W=2$.}}
            \label{chutian3}
        \end{subfigure}
\vspace{-0.5em}
       \caption{\small The OPs versus the parameters of FAS when $m=3$, $\mu=1$, and ${\gamma _{{\rm{th}}}} = 10$ dB.}
    \end{minipage}
     \vspace{-0.8em}
    \label{fig:overall}
\end{figure*}
\begin{figure*}[htbp]
    \centering
    \begin{minipage}{0.31\linewidth} 
        \centering
        \includegraphics[width=\linewidth]{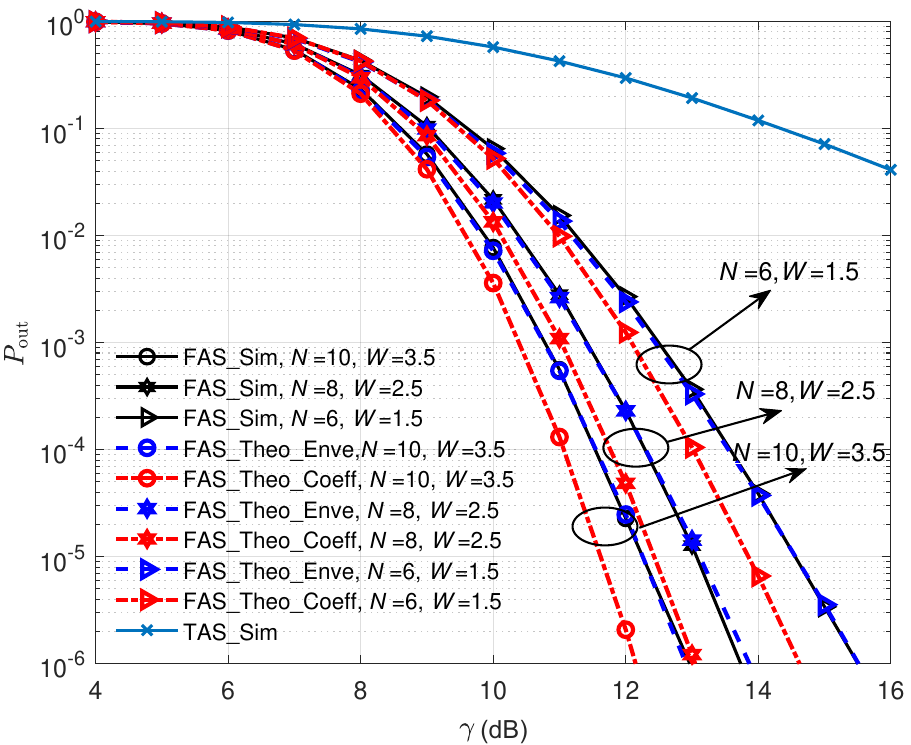}
        \caption{\small{The OPs versus the average transmit SNR $\gamma$ for selected values of $W$ and $N$ when $m=3$, $\mu=1$, and ${\gamma _{{\rm{th}}}} = 10$ dB.}}
        \label{chutian1}
    \end{minipage}
    \hfill
    \begin{minipage}{0.64\linewidth} 
        \centering
        \begin{subfigure}[b]{0.495\linewidth}
            \centering
            \includegraphics[width=\linewidth]{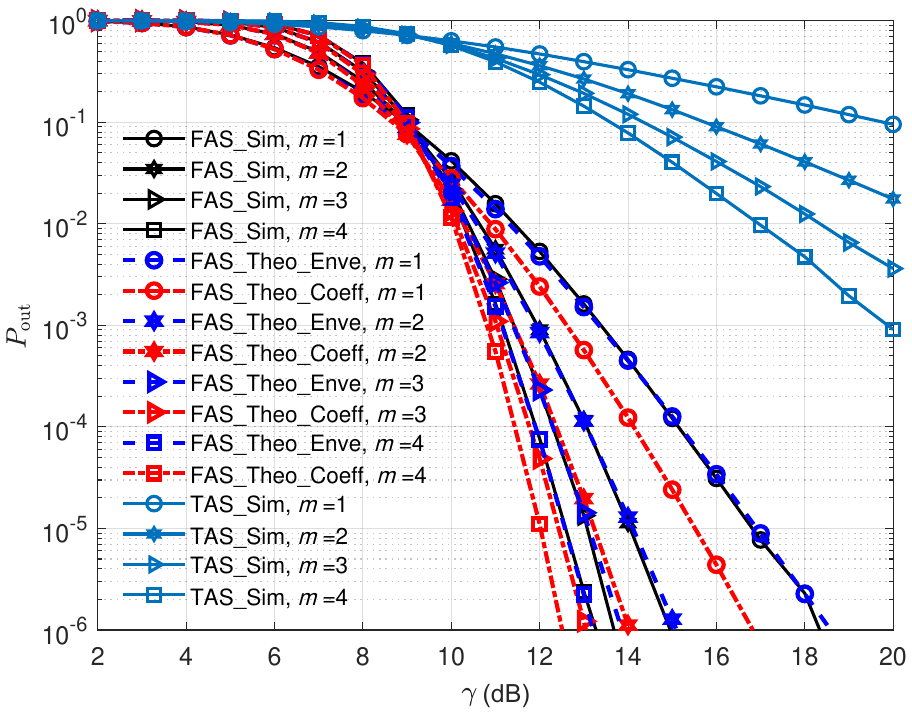}
            \caption{\small{For selected values of $m$ when $W=2.5$, $N=8$, $\mu=1$, and ${\gamma _{{\rm{th}}}} = 10$ dB.}}
            \label{chutian2}
        \end{subfigure}
        \hfill
        \begin{subfigure}[b]{0.48\linewidth}
            \centering
            \includegraphics[width=\linewidth]{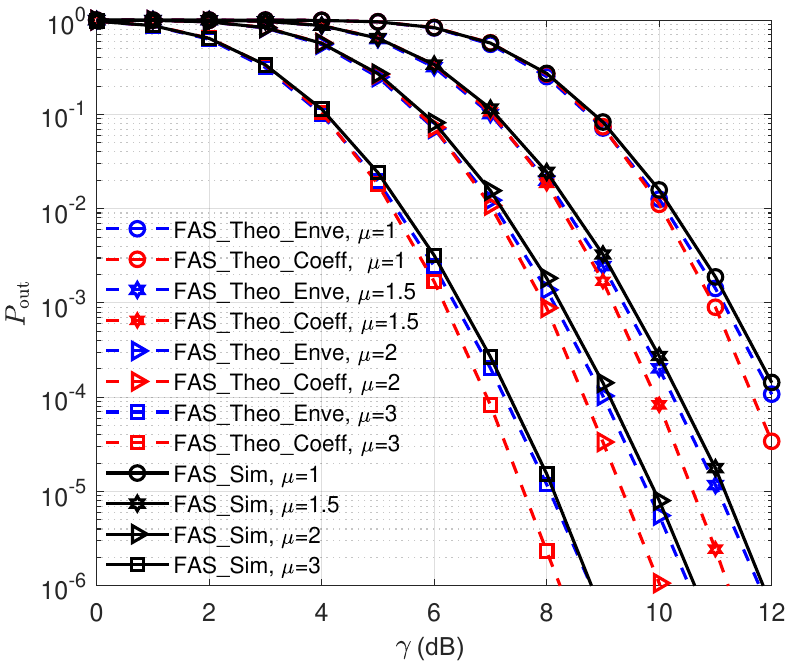}
            \caption{\small{For selected values of $\mu$ when $W=2.5$, $N=10$, $m=3$, and ${\gamma _{{\rm{th}}}} = 10$ dB.}}
            \label{chutian3}
        \end{subfigure}
\vspace{-0.5em}
       \caption{\small The OPs versus the average transmit SNR $\gamma$ for selected Nakagami-$m$ parameters.}
    \end{minipage}
    \vspace{-1em}
    \label{fig:overall}
\end{figure*}

In this section, the fully correlated Nakagami-$m$ channel envelopes are generated using the developed channel construction approach, and the OP of FAS is accurately evaluated through Monte-Carlo simulation, serving as a benchmark for validating the theoretical results.
 The OP of FAS exploiting Gaussian copula under fully correlated Nakagami-$m$ fading is computed by setting $F_{|{h^{\left( n \right)}}|}^{}(r)$ as the CDF of Nakagami-$m$ distribution and $r = \hat \gamma $ in \eqref{1A}, where $\hat \gamma  = \sqrt {{{{\gamma _{{\rm{th}}}}} \mathord{\left/
 {\vphantom {{{\gamma _{{\rm{th}}}}} \gamma }} \right.
 \kern-\nulldelimiterspace} \gamma }}$, $\gamma $ denotes the average transmit signal-to-noise ratio (SNR), and ${{\gamma _{{\rm{th}}}}}$ is the decoding SNR threshold.
In all figures, `TAS\_Sim' and `FAS\_Sim' denote the Monte Carlo simulation results of point-to-point TAS and FAS, respectively.
`FAS\_Theo\_Enve' and `FAS\_Theo\_Coeff' denote the theoretical results derived from Gaussian copula using envelope- and coefficient-level correlation matrices, respectively.

Fig. 4 depicts the OPs versus the average transmit SNR $\gamma$.
Overall, both `FAS\_Theo\_Enve' and `FAS\_Theo\_Coeff' can effectively model the outage behavior of FAS, while noticeable discrepancies remain between `FAS\_Theo\_Coeff' and `FAS\_Sim', especially in sparse port deployment and the low-OP regime, compared to the closer match observed with `FAS\_Theo\_Enve'.
Specifically, when $W$ is large (e.g., $W=3.5$, $N=10$), i.e., sparse port deployment, `FAS\_Theo\_Enve' exhibits a clear advantage in modeling accuracy in the low-OP regime.
When $W$ is small (e.g., $W=0.5$, $N=10$), i.e., dense port deployment, both `FAS\_Theo\_Coeff' and `FAS\_Theo\_Enve' yield comparable modeling accuracy.

Fig. 5 (a) shows the variation of OPs with respect to $W$ for selected values of $N$.
It is evident that `FAS\_Theo\_Enve' offers high accuracy on the whole.
In the high-$W$ regime with given $N$ (sparse port deployment), `FAS\_Theo\_Enve' have an improvement in accuracy compared to `FAS\_Theo\_Coeff'.
Moreover, the OPs exhibit fluctuations in the high $W$ regime. This behavior stems from the oscillatory nature of ${J_0}\left(  \cdot  \right)$ in the fully correlated model, where the channel at each port is constructed as a weighted combination of that at other ports.
Fig. 5 (b) depicts the OPs versus $N$ for selected values of $\gamma$. The matching trend further confirms that `FAS\_Theo\_Enve' obtains more accurate OP than `FAS\_Theo\_Coeff' in sparse port deployment, i.e., in the low-$N$ regime with given $W$. As $\gamma$ increases, `FAS\_Theo\_Enve' yields more accurate results than `FAS\_Theo\_Coeff' under denser port deployment, i.e., lager $N$ given a fixed $W$.

\begin{table}
\centering
\renewcommand{\arraystretch}{1.1} 
\caption{\small{The correlated coefficients of the generated channel envelopes under spare port deployment with $m=3$}}
\begin{adjustbox}{max width=\textwidth}
\begin{tabular}{c@{\hspace{2pt}}||@{\hspace{2pt}}c@{\hspace{2pt}}|@{\hspace{2pt}}c@{\hspace{2pt}}|@{\hspace{2pt}}c@{\hspace{2pt}}||@{\hspace{2pt}}c@{\hspace{2pt}}|@{\hspace{2pt}}c@{\hspace{2pt}}|@{\hspace{2pt}}c@{\hspace{2pt}}}
\hline
\multirow{2}{*}{} & \multicolumn{3}{c@{\hspace{2pt}}||@{\hspace{2pt}}}{$N=10, W=3.5$} & \multicolumn{3}{c@{\hspace{2pt}}}{$N=8, W=2.5$} \\
\cline{2-7}
                    & Sim & Coeff & Enve & Sim & Coeff & Enve \\
\hline
${{\bf{J}}_{1,2}}$ & -0.0159   & -0.0243  & -0.0093  & 0.0025   & 0.0779    & -0.0123 \\
${{\bf{J}}_{1,3}}$ & 0.0943    & 0.2796   & 0.0633   & 0.0795   & 0.2786    & 0.0886  \\
${{\bf{J}}_{1,4}}$ & 0.0424    & -0.2193  & 0.0624   & 0.0133   & -0.0953   & 0.0016  \\
${{\bf{J}}_{1,5}}$ & 0.0254    & 0.1061   & 0.0195   & 0.0221   & -0.0998   & 0.0104  \\
${{\bf{J}}_{1,6}}$ & -0.0077   & 0.0426   & 0.0158   & 0.0506   & 0.2157    & 0.0414  \\
${{\bf{J}}_{1,7}}$ & 0.0368    & -0.1551  & 0.0246   & 0.0319   & -0.1518   & 0.0105  \\
${{\bf{J}}_{1,8}}$ & 0.0356    & 0.1750   & 0.0369   &         &           &         \\
${{\bf{J}}_{1,9}}$ & 0.0224    & -0.1353  & 0.0175   &         &           &         \\
\hline
\end{tabular}
\end{adjustbox}
\vspace{-10pt}
\end{table}

Under sparse port deployment, Fig. 6 shows the OPs versus $\gamma$ under different methods, for selected values of $W$ and $N$.
The discrepancies between `FAS\_Sim' and `FAS\_Theo\_Coeff' as well as `FAS\_Theo\_Coeff' further validate the above conclusion.
Notably, the OPs obtained from `FAS\_Theo\_Coeff' consistently lie below the accurate OPs obtained from `FAS\_Sim', indicating that the Gaussian copula using coefficient-level correlation matrix in existing works underestimates the outage performance of FAS in sparse port deployment.
To explain this phenomenon,
Table I compares the resulting channel envelope correlation coefficients (due to the limitation of space, only the channel envelope correlation coefficients between the first port and other ports are presented) obtained from Gaussian copula using the coefficient- and envelope-level correlation matrices.
It can be observed that the correlation coefficients of the generated channel envelopes based on the envelope-level correlation matrix are much closer to those from Monte-Carlo simulation.
However, the coefficient-level correlation matrix produces multiple negative correlation coefficients of the generated channel envelopes.
This reduces the probability that multiple channels experience deep fading simultaneously, thereby underestimating the true outage performance of FAS, since the port with the best channel condition is typically selected for transmission.
Moreover, low-probability outage events are mainly caused by deep fading of the channels, and thus the OP in the low-outage regime is particularly sensitive to simultaneous deep fading across multiple channels.
Therefore, employing the Gaussian copula based on the coefficient-level correlation matrix introduces noticeable modeling errors in the low-outage regime.
Besides, as $\gamma$ increases, $\hat \gamma$ decreases, and thus the outage events are mainly caused by deep fading of the channels. Therefore, noticeable modeling errors exists for high $\gamma$.


To further explore the impact of Nakagami-$m$ fading parameters on the modeling accuracy of the Gaussian copula when different correlation matrices are used, Fig. 7 plots the OPs versus the average transmit SNR $\gamma$ under different methods, for selected values of $m$ and $\mu$, respectively.
On the whole, regardless of the values of $m$ and $\mu$, `FAS\_Theo\_Enve' consistently provides more accurate OP performance than `FAS\_Theo\_Coeff' in sparse port deployment and the low OP regime.
Moreover, as $m$ decreases or $\mu$ increases, `FAS\_Theo\_Enve' achieves higher accuracy than `FAS\_Theo\_Coeff'.

\vspace{-0.5em}

\section{Conclusions}

This paper explored the modeling accuracy of using envelope-level correlation matrix under Gaussian copula framework, and developed a method to generate fully correlated Nakagami-$m$ fading channel envelopes, enabling exact Monte-Carlo simulations for validating the theoretical results.
Simulation results confirm that using the envelope-level correlation matrix yields higher modeling accuracy, especially in sparse port deployment and low-outage regime.
Moreover, as the average transmit SNR increases, Gaussian copula using the envelope-level correlation matrix yields
more accurate OP than using coefficient-level one under denser port deployment.
As the Nakagami-$m$ fading parameters $m$ decreases or
$\mu$ increases, envelope correlation-based Gaussian copula achieves higher accuracy than the coefficient correlation matrix. Building on these, in ultra-reliable communication or sparse port deployment scenarios, Gaussian copula using the envelope-level correlation matrix provides an effective performance evaluation approach.
Moreover, it can be further validated under other fading channel models, such as cascaded channels.

\vspace{-0.5em}

\ifCLASSOPTIONcaptionsoff
  \newpage
\fi
\bibliographystyle{IEEEtran}
\bibliography{refa}

\begin{thebibliography}{1}
\providecommand{\url}[1]{#1}
\csname url@samestyle\endcsname
\providecommand{\newblock}{\relax}
\providecommand{\bibinfo}[2]{#2}
\providecommand{\BIBentrySTDinterwordspacing}{\spaceskip=0pt\relax}
\providecommand{\BIBentryALTinterwordstretchfactor}{4}
\providecommand{\BIBentryALTinterwordspacing}{\spaceskip=\fontdimen2\font plus
\BIBentryALTinterwordstretchfactor\fontdimen3\font minus
  \fontdimen4\font\relax}
\providecommand{\BIBforeignlanguage}[2]{{%
\expandafter\ifx\csname l@#1\endcsname\relax
\typeout{** WARNING: IEEEtran.bst: No hyphenation pattern has been}%
\typeout{** loaded for the language `#1'. Using the pattern for}%
\typeout{** the default language instead.}%
\else
\language=\csname l@#1\endcsname
\fi
#2}}
\providecommand{\BIBdecl}{\relax}
\BIBdecl

\bibitem{10678877}
F.~R. Ghadi, K.-K. Wong \emph{et~al.}, ``A {Gaussian} copula approach to the
  performance analysis of fluid antenna systems,'' \emph{IEEE Trans. Wireless
  Commun.}, vol.~23, no.~11, pp. 17\,573--17\,585, 2024.

\bibitem{10753482}
W.~K. New, K.-K. Wong \emph{et~al.}, ``A tutorial on fluid antenna system for
  {6G} networks: Encompassing communication theory, optimization methods and
  hardware designs,'' \emph{IEEE Commun. Surveys Tuts.}, vol.~27, no.~4, pp.
  2325--2377, 2025.

\bibitem{9264694}
K.-K. Wong, A.~Shojaeifard \emph{et~al.}, ``Fluid antenna systems,'' \emph{IEEE
  Trans. Wireless Commun.}, vol.~20, no.~3, pp. 1950--1962, 2021.

\bibitem{wong2022closed}
K.~Wong, K.~Tong, Y.~Chen, and Y.~Zhang, ``Closed-form expressions for spatial
  correlation parameters for performance analysis of fluid antenna systems,''
  \emph{Electron. Lett.}, vol.~58, no.~11, pp. 454--457, 2022.

\bibitem{10623405}
P.~Ram{\'\i}rez-Espinosa, D.~Morales-Jimenez, and K.-K. Wong, ``A new spatial
  block-correlation model for fluid antenna systems,'' \emph{IEEE Trans.
  Wireless Commun.}, vol.~23, no.~11, pp. 15\,829--15\,843, 2024.

\bibitem{10103838}
M.~Khammassi, A.~Kammoun, and M.-S. Alouini, ``A new analytical approximation
  of the fluid antenna system channel,'' \emph{IEEE Trans. Wireless Commun.},
  vol.~22, no.~12, pp. 8843--8858, 2023.

\bibitem{9159617}
F.~R. Ghadi and G.~A. Hodtani, ``Copula-based analysis of physical layer
  security performances over correlated rayleigh fading channels,'' \emph{IEEE
  Trans. Inf. Forensics Security}, vol.~16, pp. 431--440, 2021.

\bibitem{1356206}
Y.~Chen and C.~Tellambura, ``Distribution functions of selection combiner
  output in equally correlated {Rayleigh, Rician, and Nakagami-m} fading
  channels,'' \emph{IEEE Trans. Commun.}, vol.~52, no.~11, pp. 1948--1956,
  2004.

\bibitem{stuber2001principles}
G.~L. St{\"u}ber, \emph{Principles of mobile communication}, 2nd ed. Norwell,
  MA, USA: Kluwer, 2002.

\end{thebibliography}

\end{document}